\documentclass[12pt,axodraw]{article}
\input psfig.sty
\input axodraw.sty

\def\Re{{\cal R \mskip-4mu \lower.1ex \hbox{\it e}\,}}
\def\Im{{\cal I \mskip-5mu \lower.1ex \hbox{\it m}\,}}
\def\ie{{\it i.e.}}

\def\etc{{\it etc}}
\def\etal{{\it et al.}}

\def\sub#1{_{\lower.25ex\hbox{$\scriptstyle#1$}}}
\def\tev{\,{\ifmmode\mathrm {TeV}\else TeV\fi}}
\def\gev{\,{\ifmmode\mathrm {GeV}\else GeV\fi}}
\def\mev{\,{\ifmmode\mathrm {MeV}\else MeV\fi}}
\def\mpl{\ifmmode \overline M_{Pl}\else $\overline M_{Pl}$\fi}
\def\to{\rightarrow}

\def\subw{_{\rm w}}
\def\mh{\ifmmode m\sbl H \else $m\sbl H$\fi}
\def\mch{\ifmmode m_{H^\pm} \else $m_{H^\pm}$\fi}
\def\mt{\ifmmode m_t\else $m_t$\fi}
\def\mc{\ifmmode m_c\else $m_c$\fi}
\def\mz{\ifmmode M_Z\else $M_Z$\fi}
\def\mw{\ifmmode M_W\else $M_W$\fi}
\def\mws{\ifmmode M_W^2 \else $M_W^2$\fi}
\def\mhs{\ifmmode m_H^2 \else $m_H^2$\fi}   
\def\mzs{\ifmmode M_Z^2 \else $M_Z^2$\fi}
\def\mts{\ifmmode m_t^2 \else $m_t^2$\fi}
\def\mcs{\ifmmode m_c^2 \else $m_c^2$\fi}
\def\mchs{\ifmmode m_{H^\pm}^2 \else $m_{H^\pm}^2$\fi}
\def\ztwo{\ifmmode Z_2\else $Z_2$\fi}
\def\zone{\ifmmode Z_1\else $Z_1$\fi}
\def\mtwo{\ifmmode M_2\else $M_2$\fi}
\def\mone{\ifmmode M_1\else $M_1$\fi}
\def\tb{\ifmmode \tan\beta \else $\tan\beta$\fi}
\def\xw{\ifmmode x\subw\else $x\subw$\fi}
\def\ch{\ifmmode H^\pm \else $H^\pm$\fi}
\def\lum{\ifmmode {\cal L}\else ${\cal L}$\fi}
\def\inpb{\,{\ifmmode {\mathrm {pb}}^{-1}\else ${\mathrm {pb}}^{-1}$\fi}}
\def\infb{\,{\ifmmode {\mathrm {fb}}^{-1}\else ${\mathrm {fb}}^{-1}$\fi}}
\def\epem{\ifmmode e^+e^-\else $e^+e^-$\fi}
\def\ppb{\ifmmode \bar pp\else $\bar pp$\fi}
\def\bsg{\ifmmode B\to X_s\gamma\else $B\to X_s\gamma$\fi}
\def\bsll{\ifmmode B\to X_s\ell^+\ell^-\else $B\to X_s\ell^+\ell^-$\fi}
\def\bstt{\ifmmode B\to X_s\tau^+\tau^-\else $B\to X_s\tau^+\tau^-$\fi}
\def\lamt{\ifmmode \tilde\lambda\else $\tilde\lambda$\fi}
\def\shat{\ifmmode \hat s\else $\hat s$\fi}
\def\that{\ifmmode \hat t\else $\hat t$\fi}
\def\uhat{\ifmmode \hat u\else $\hat u$\fi}

\newskip\zatskip \zatskip=0pt plus0pt minus0pt
\def\matth{\mathsurround=0pt}

\def\atversim#1#2{\lower0.7ex\vbox{\baselineskip\zatskip\lineskip\zatskip
  \lineskiplimit 0pt\ialign{$\matth#1\hfil##\hfil$\crcr#2\crcr\sim\crcr}}}

\renewcommand{\thefootnote}{\fnsymbol{footnote}}

\begin{document} \begin{titlepage} 
\rightline{\vbox{\halign{&#\hfil\cr
&SLAC-PUB-9043\cr
&November 2001\cr}}}
\begin{center}

{\Large\bf Non-Commutativity and Unitarity Violation in Gauge Boson Scattering}
\footnote{Work supported by the Department of 
Energy, Contract DE-AC03-76SF00515}
\medskip

\normalsize 
{\bf \large J.L. Hewett, F.J. Petriello and T.G. Rizzo}
\vskip .3cm
Stanford Linear Accelerator Center \\
Stanford University \\
Stanford CA 94309, USA\\
\vskip .2cm

\end{center}

\begin{abstract} 

We examine the unitarity properties of spontaneously broken non-commutative
gauge theories.  We find that the symmetry breaking mechanism in the 
non-commutative Standard Model of Chaichian \etal\ leads to an unavoidable 
violation of tree-level unitarity in gauge boson scattering at high energies.  
We then study a variety of simplified spontaneously broken non-commutative 
theories and isolate the source of this unitarity violation.  Given the 
group theoretic restrictions endemic to non-commutative model building, 
we conclude that it is difficult to build a non-commutative Standard Model 
under the Weyl-Moyal approach that preserves unitarity.

\end{abstract}

\renewcommand{\thefootnote}{\arabic{footnote}} \end{titlepage}


\section{Introduction} 

The possibility of non-commuting space-time coordinates is an intriguing
one which arises naturally in string theory and gives rise to a rich
phenomenology\cite{ncrev,hist}.  As such, noncommutative quantum field
theory (NCQFT) has the potential to provide an attractive and motivated
theory of physics beyond the Standard Model (SM).  However,
while a consistent noncommutative (NC) version of QED has been 
developed, a NC analog of the full SM
has proven to be problematic for reasons detailed below.  Possible
NC gauge groups and matter representations are theoretically restricted,
and as we will
show, the basic difficulty lies in constructing a viable symmetry breaking
mechanism which reduces the NC gauge group to that of the
SM.  We investigate a variety of models and demonstrate that the NC
symmetry breaking mechanism in a leading candidate for a 
NCSM\cite{Chaichian:2001py} leads to an unavoidable 
violation of tree-level unitarity in gauge boson scattering.

NCQFT can be realized\cite{strings} in the string theoretic limit where
string endpoints propagate
in the presence of background fields.  In this case, the endpoints of
strings no longer commute and their coordinate operators obey the relation
\begin{equation}
[\hat x_\mu, \hat x_\nu] = i\theta_{\mu\nu} = {i\over\Lambda^2_{NC}}
c_{\mu\nu}\,,
\end{equation}
where $\theta_{\mu\nu} (c_{\mu\nu})$ is a constant, real, anti-symmetric 
matrix
and $\Lambda_{NC}$ represents the scale where NC effects set in.  The
most likely value for the NC scale is that of the string or fundamental
Planck scale which, in principle, can be of order a TeV.  Space-space
noncommutativity occurs in the presence of background magnetic fields
with $\theta_{ij}\equiv(\hat c_B)_{ij}/\Lambda_B^2\neq 0$, 
whereas space-time NC
theories are related to background electric fields with $\theta_{0i}
\equiv(\hat c_E)_{0i}/\Lambda_E^2\neq 0$.  Note that the unit vectors $\hat 
c_{B,E}$ are frame independent and hence Lorentz invariance will be
violated at the $\Lambda_{B,E}$ NC scales\cite{lorviol}.  
These theories conserve
CPT and have been claimed to be unitary if the condition $\theta_{\mu\nu}
\theta^{\mu\nu}\geq 0$ is satisfied\cite{unitary}.  While this condition
might hold for unbroken NC theories, we will show 
below that it is not sufficient to guarantee unitarity for all choices
of Higgs representations in the spontaneously broken case.

There are two approaches for constructing quantum field theories on NC 
spaces.  The first relates fields in the NC and the usual $R^4$ spaces via
the Weyl-Moyal correspondence.  In this case, ordinary fields are replaced
by their NC counterparts and the ordering of fields
in the NC space is given by the Moyal star product, defined by
\begin{eqnarray}
\hat A(\hat x)\hat B(\hat x) & = & A(x)*B(x)\\
& = & A(x) \exp\{{i\over 2}\theta^{\mu\nu}\overleftarrow\partial_\mu
\overrightarrow\partial_\nu\}
B(x) \,,\nonumber
\end{eqnarray}
where the hatted quantities correspond to those
in the NC space.  NCQFT is formulated as that of ordinary QFT with star 
products substituting for products and commutators replaced by Moyal
brackets,
\begin{equation}
[A,B]_{MB} = A*B - B*A\,,
\end{equation}
with $\int d^4x[A(x),B(x)]_{MB}=0$. Only $U(n)$ Lie algebras are closed 
under Moyal brackets\cite{matsubara} and hence NC model building is
restricted in this case.  The simplest version of a NCSM thus requires
a gauge group at least as large as $U(3)\times U(2)\times U(1)$.

The second approach relies on the Seiberg-Witten map which 
represents the NC fields 
in terms of the corresponding fields in ordinary $R^4$ as 
a power series expansion in $\theta_{\mu\nu}$.  While this procedure
allows for generalizations of the star product to be constructed for
a broader set of gauge theories\cite{wess}, it generates an 
infinite set of higher dimensional operators.  It is hence difficult 
to employ and is not necessarily attractive as a model building option.  
We neglect this possibility here.

NC extensions have been shown to preserve renormalizability at the
one- to two-loop level in the cases of unbroken\cite{renorm1} 
$\phi^4$ and Yang-Mills theories, as well as for spontaneously
broken\cite{renorm2} $U(1)$ and $U(2)$ at one-loop.  

Ordinary gauge transformations for $U(n)$ gauge theories must be
modified\cite{haya} to include NC generalizations.  Gauge invariance
requires non-abelian like gauge couplings, even for $U(1)$ theories, and
dictates that matter only be placed in the singlet, (anti-)fundamental,
or adjoint representations.  
In addition, some interaction vertices pick up momentum
dependent phase factors.  In the NC
version of QED, this induces 3- and 4-point self-couplings for 
photons and restricts QED interactions to particles of charge $q=0,
\pm e$.  Despite these limitations, $2\to 2$ scattering processes at 
high energies in QED provide clean observables\cite{pheno} for NCQED 
effects.  In addition, several low-energy QED processes\cite{lowe}, 
{\it e.g.,} Lamb shift, $g-2$, can probe NC interactions. 

There are clear difficulties in formulating a NCSM based on the 
Weyl-Moyal approach due to the restrictions imposed from group
theory and gauge invariance: (i) charge quantization, {\it e.g.}, NCQED
cannot contain fractionally charged particles, and (ii) performing the NC
$U(n)$ symmetry breaking without generating
a mass for the gauge fields of the $SU(n)$ component of the $U(n)$.
A first attempt to build a 
NCSM by Chaichian \etal\cite{Chaichian:2001py} is based on the gauge 
group $U(3)_c\times U(2)_L\times U(1)$ and at first appears to resolve
these issues.  However, we find that the model
proposed by these authors leads to unitarity violation at high
energies in $2\to 2$ gauge boson scattering.  By studying similar
scattering processes in more simplified NC gauge theories, we
have isolated the cause of the unitarity breakdown in this model to
be the choice of NC symmetry breaking mechanism.  

A review of the
Chaichian \etal\ model and our calculations of gauge boson scattering
within it is given in the next section.  Our study of potential 
unitarity violation in $2\to 2$ gauge scattering in a variety of
simpler NC models is presented in Section III, and a discussion
of our results is given in Section IV.

\section{The Non-commutative Standard Model}

As was discussed in the introduction, group theory and gauge invariance 
enforce strict constraints on the construction of NC models following the 
conventional Moyal approach. In addition to 
the requirement that theories be built out of products of $U(n)$ factors and 
that only the singlet ($S$), fundamental ($F$), anti-fundamental ($\bar F$) or 
adjoint ($A$) representations are allowed for matter fields (\ie, $Q=0, \pm 1$ 
for the case of $U(1)$), Chaichian \etal ~have 
shown~\cite{Chaichian:2001mu}  that a ``No Go" theorem is operative. This 
theorem states that for any gauge group consisting of a product of simple 
group factors, 
matter fields can only transform nontrivially under at most two of 
the simple groups. In particular, if a field transforms under one factor as a 
$F$ then it must transform under the second as an $\bar F$. 
Note that while SM matter fields are already in fundamental 
representations, the left-handed quark doublet($Q_L$) transforms nontrivially 
under all {\it three} SM group factors.  This clearly complicates the
construction of the NCSM.

The minimal group structure for the NCSM which can  
contain $SU(3)_c\times SU(2)_L\times U(1)_Y$ in the commutative limit is 
$U(3)_c\times U(2)_L\times U(1)$; this has both positive and negative
features for NCSM construction. On the positive side, the SM fractional
hypercharges may receive contributions arising from the $U(1)$ 
factor as well as from the two $U(1)$ 
subgroups, $U(1)_{c,L}$, contained in $U(3)_c$ and $U(2)_L$ 
so that the constraint of $U(1)$ charge
$Q=0, \pm 1$ assignments can be satisfied. 
However, the increase in group rank by two implies 
that there are now two new additional neutral weak bosons in the theory that 
will have couplings to SM matter fields. These two new states must be made 
sufficiently massive as to avoid present Tevatron direct search 
constraints~\cite{Tevatron} as well as those arising from precision 
electroweak data~\cite{Precision}; we thus expect their masses to be greater 
than a TeV or 
so. In addition, spontaneous symmetry breaking must take 
place in at least two steps to recover the correct phenomenological structure 
and have the correct SM commutative limit. Effectively, this means that 
the $U(1)\times U(1)_c \times U(1)_L$ symmetry must first break 
to $U(1)_Y$ {\it without} breaking the $SU(3)_c$ or $SU(2)_L$ 
groups themselves.

\vspace{0.5cm}
\begin{table*}[htpb]
\begin{center}
\label{reps}
\begin{tabular}{lccc}
\hline
\hline
Field & $U(3)$ & $U(2)$ & $U(1)$\\
\hline
$Q_L$ & $\bar F$ & $F$ &$S$   \\
$L_L$ & $S$  &$F$  &$\bar F$ \\ 
$e_R$ & $S$  &$S$  &$\bar F$ \\
$u_R$ & $\bar F$ &$S$& $F$   \\
$d_R$ & $\bar F$ &$S$& $S$   \\
$h$   & $S$ & $F$ & $S$      \\
\hline
\hline
\end{tabular}
\caption{Quantum number assignments for the SM fermion and Higgs fields in the 
NCSM using the notation discussed in the text.}
\end{center}
\end{table*}

In a recent paper, Chaichian~\cite{Chaichian:2001py} have apparently either 
overcome or satisfied all of the above obstacles and constructed a NC version 
of the SM. The SM matter content of this proposed construction is given in 
Table 1. As can be seen the representation content explicitly satisfies the 
No-Go theorem. Next, the authors perform the breaking of the
$U(3)_c\times U(2)_L \times U(1)$ symmetry in several steps: first, two linear 
combinations of the $U(1)$'s within $U(1)_c \times U(1)_L \times U(1)$ must 
be broken. As this is 
a product of three group factors the No-Go theorem requires that two Higgs 
fields are necessary to break this symmetry down to $U(1)_Y$. It is clear that 
these Higgs fields cannot correspond to any of the usual $SU(n)$ 
representations since  then
their vev's would not only break the $U(1)$ subgroups of 
the $U(n)$'s but the $SU(n)$ factors as well.  To get 
around this, Chaichian \etal~\cite{Chaichian:2001py} have 
considered Higgs fields in 
a new representation, which they call Higgsac ($H$) fields.  They transform
only under the $U(1)$ subgroup of $U(n)$, so that when they acquire a
vev the $SU(n)$ subgroup remains unbroken.
Essentially, a Higgsac breaks $U(1)_c \times U(1)_L$ to $U(1)'$ and 
a second Higgsac then produces the breaking $U(1)' \times U(1)$ to $U(1)_Y$. 
Given the two distinct Higgsac vev's, the masses of the two heavy gauge bosons 
are {\it uncorrelated} and arbitrary. To be more specific, denoting the 
original weak eigenstate $U(1)_c \times U(1)_L \times U(1)$ gauge fields 
by $G^{0'},W^{0'}$ and $B$, respectively, the weak
and the unprimed mass eigenstate fields are related by:
$G^{0'}=c_{23}G^0+s_{23}(c_{11}W^0+s_{11}Y)$, 
$W^{0'}=-s_{23}G^0+c_{23}(c_{11}W^0+s_{11}Y)$, and $B=-s_{11}W^0+c_{11}Y$. 
Here $W^0,G^0$ are the new massive gauge bosons while $Y$ is the massless boson 
coupling to hypercharge and $c_{ij}(s_{ij})=
\cos \delta_{ij}(\sin \delta_{ij})$, with $\delta_{ij}$ being appropriate 
mixing angles. As mentioned above we anticipate 
that these new gauge bosons are more massive than about $\sim 1$ TeV. Given 
the gauge couplings $g_{1,2,3}$ for the appropriate $U(n)$ groups, the 
mixing angles can be expressed as $\tan \delta_{23}=2g_2/3g_3$ and 
$2c_{23} \tan \delta_{11}=g_1/g_2$. In addition, consistency with the fermion 
couplings of the SM in the commutative limit implies that $g_3=g_s$, the 
usual QCD coupling, $g_2=g$, the usual weak coupling, $g_1=2g'/c_{11}$, 
with $g'$ being the usual hypercharge coupling of the SM, and also the relation 
$s_{11}c_{23}=\tan \theta_W$. 

The last step of the symmetry breaking is accomplished through the vev of the 
isodoublet field $h$ listed 
in the Table. It generates the ordinary fermion masses 
in the usual manner as well as the conventional $W^\pm$ and $Z$ masses with the 
identification $e=g \sin \theta_W$. There is however one additional effect: 
mixing is induced between the SM $Z$ field and the more massive $G^0,W^0$ 
states which is of order $m_Z^2/m_{G,W}^2$; for TeV or heavier states we 
expect this mixing to be quite small, $\leq 0.01$, and can be 
safely neglected on most occasions. 
Given the values of the couplings, all of the mixing angles are fixed so that 
the only free parameters in the gauge sector of the model are the masses of the 
$G^0$ and $W^0$ states. Of course the scales $\Lambda_{E,B}$ associated 
with the NC physics, as defined in the introduction, still remain arbitrary.

Before examining the phenomenology of this model, 
we may first want to check whether or not 
the theory is unitary at tree-level.  The classic 
test for this in the SM is the scattering of pairs of 
longitudinal $W^\pm$, \ie, 
$W^+_L W^-_L \to W^+_L W^-_L$. This process provides a particularly useful 
test in the NCSM 
case as it is independent of how the fermions are embedded into the theory. 
In the SM at amplitude level, the leading terms from a typical diagram behave 
as $(s^2/m_W^4)$; such terms cancel among the contributions from the 
$s$- and $t$-channel $\gamma$ and 
$Z$ exchanges and the 4-point graph due to gauge invariance and the natural 
relationship $m_W=m_Z \cos \theta_W$. The sub-leading $s/m_W^2$ contributions 
also cancel when $s$- and $t$-channel Higgs boson exchanges are included 
yielding a result which does not grow with $s$ and 
is unitary. In the NCSM, as we will see below, the leading terms still 
cancel because of the gauge symmetry, but the subleading terms 
remain so that unitarity is not satisfied. Let us now discuss these 
cancellations  in some detail. 

To be as explicit and straightforward as possible, we will separately 
calculate both the leading and next-to-leading $s/m_W^2$ terms for each of 
the contributing diagrams to demonstrate how the required cancellations fail 
to occur in the NCSM.  For simplicity, these 
calculations will be performed in the limit that the weak scale mixing induced 
between the $Z$ and the heavy states $W^0$ and $G^0$ can be neglected. The 
additional terms that are generated by such mixing are subleading to the ones 
included below by factors of order $m_Z^2/m_{G,W}^2$ and will have no 
influence upon our results. 

\begin{figure}
\begin{picture}(350,585)(10,10)
\put(10,56){$h$}
\Line(10,50)(50,50)
\Photon(50,50)(78.28,78.28){5}{3}
\put(64,55){$\nearrow$ $p_1$}
\put(82,74){$W^{+}_{\mu}$}
\Photon(50,50)(78.28,21.72){5}{3}
\put(80,20){$W^{-}_{\nu}$}
\put(64,41){$\searrow$ $p_2$}
\put(100,50){$=ig \, m_W e^{i\, p_1 \wedge p_2} g_{\mu \nu}$}

\Photon(10,148.28)(50,148.28){5}{3}
\put(10,158){$A_{\rho}$}
\put(10,135){$\leftarrow$  $p_3$}
\Photon(50,148.28)(78.28,176.56){5}{3}
\put(64,153.28){$\nearrow$ $p_1$}
\put(82,168.28){$W^{-}_{\mu}$}
\Photon(50,148.28)(78.28,120){5}{3}
\put(80,118){$W^{+}_{\nu}$}
\put(64,139){$\searrow$ $p_2$}
\put(100,148.28){$= -g \left\{ is_{W} \, {\rm cos}\left(p_1 \wedge p_2 \right)
+c_{23}s_{11}c_{W} \, {\rm sin}\left( p_1 \wedge p_2 \right) \right\} 
 \bigg[ \, (p_1 -p_2)_{\rho} g_{\mu\nu} $}
\put(134,123.28){$ +(p_2 -p_3)_{\mu}g_{\nu \rho} 
+(p_3 -p_1)_{\nu}g_{\mu\rho}\, \bigg]$}

\Photon(10,246.56)(50,246.56){5}{3}
\put(10,256.28){$Z_{\rho}$}
\put(10,233.28){$\leftarrow$  $p_3$}
\Photon(50,246.56)(78.28,274.84){5}{3}
\put(64,251.56){$\nearrow$ $p_1$}
\put(82,266.56){$W^{-}_{\mu}$}
\Photon(50,246.56)(78.28,218.28){5}{3}
\put(80,216.28){$W^{+}_{\nu}$}
\put(64,237.28){$\searrow$ $p_2$}
\put(100,246.56){$=  -g \left\{ ic_{W} \, {\rm cos}\left(p_1 \wedge p_2 \right)
-c_{23}s_{11}s_{W} \, {\rm sin}\left( p_1 \wedge p_2 \right) \right\} 
 \bigg[ \, (p_1 -p_2)_{\rho} g_{\mu\nu}  $}
\put(134,221.56) {$+(p_2 -p_3)_{\mu}g_{\nu \rho} 
+(p_3 -p_1)_{\nu}g_{\mu\rho} \, \bigg]$}

\Photon(10,344.84)(50,344.84){5}{3}
\put(10,354.56){$G^{0}_{\rho}$}
\put(10,331.56){$\leftarrow$  $p_3$}
\Photon(50,344.84)(78.28,373.12){5}{3}
\put(64,349.84){$\nearrow$ $p_1$}
\put(82,364.84){$W^{-}_{\mu}$}
\Photon(50,344.84)(78.28,316.56){5}{3}
\put(80,314.56){$W^{+}_{\nu}$}
\put(64,334.56){$\searrow$ $p_2$}
\put(100,344.84){$=  g \, s_{23} \, {\rm sin}\left(p_1 \wedge p_2 \right)
\left[ (p_1 -p_2)_{\rho} g_{\mu\nu} +(p_2 -p_3)_{\mu}g_{\nu \rho} 
+(p_3 -p_1)_{\nu}g_{\mu\rho}\right]$}

\Photon(10,443.12)(50,443.12){5}{3}
\put(10,452.84){$W^{0}_{\rho}$}
\put(10,429.84){$\leftarrow$  $p_3$}
\Photon(50,443.12)(78.28,471.4){5}{3}
\put(64,448.12){$\nearrow$ $p_1$}
\put(82,463.12){$W^{-}_{\mu}$}
\Photon(50,443.12)(78.28,414.84){5}{3}
\put(80,412.84){$W^{+}_{\nu}$}
\put(64,432.84){$\searrow$ $p_2$}
\put(100,443.12){$= -g \, c_{23}c_{11} \, {\rm sin}\left(p_1 \wedge p_2 \right)
\left[ (p_1 -p_2)_{\rho} g_{\mu\nu} +(p_2 -p_3)_{\mu}g_{\nu \rho} 
+(p_3 -p_1)_{\nu}g_{\mu\rho}\right]$}

\Photon(31.72,573.12)(60,544.84){5}{3}
\put(20,549.84){$p_1$ $\nwarrow$}
\put(10,564.84){$W^{+}_{\mu}$}
\Photon(31.72,516.56)(60,544.84){5}{3}
\put(20,535.56){$p_3$ $\swarrow$}
\put(10,514.56){$W^{+}_{\nu}$}
\Photon(60,544.84)(88.28,573.12){5}{3}
\put(74,549.84){$\nearrow$ $p_4$}
\put(92,564.84){$W^{-}_{\sigma}$}
\Photon(60,544.84)(88.28,516.56){5}{3}
\put(90,514.56){$W^{-}_{\rho}$}
\put(74,535.56){$\searrow$ $p_2$}
\put(110,544.84){$= ig^2 \, \bigg\{ \left[{\rm cos}\left(p_1 \wedge p_2 +
p_3 \wedge p_4 \right) +{\rm cos}\left(p_1 \wedge p_4 +
p_2 \wedge p_3 \right) \right]g_{\mu\nu}g_{\rho\sigma}$}
\put(143,523.28){$-{\rm cos}\left( p_1 \wedge p_4 +p_2 \wedge p_3 \right) 
g_{\mu\rho}g_{\nu\sigma} $}
\put(143,501.72){$-{\rm cos}\left( p_1 \wedge p_2 +p_3 \wedge p_4 \right) 
g_{\mu\sigma}g_{\nu\rho} \, \bigg\} $}
\end{picture}
\caption{NCSM Feynman rules relevant for $W^{+}_L W^{-}_L 
\rightarrow W^{+}_L W^{-}_L$.}
\label{ncsmfeyn}
\end{figure}

Using the couplings derived in \cite{renorm1} for NC $U(n)$ gauge
theories, we can now generate all of the Feynman rules for this model.
The relevant ones for $W_L^+W_L^-\to W_L^+W_L^-$ are presented in
Fig. \ref{ncsmfeyn}.  Here, we have introduced the wedge product
which is defined as $p_i\wedge p_j={1\over 2}p_i^\mu p_j^\nu\theta_{\mu\nu}$,
and the direction of the momenta are as labeled.
The kinematic phases at
each vertex, which arise from the Fourier transformation of the
interaction term into momentum space, mentioned above are explicitly 
apparent.  To be specific, we will perform our calculation in the 
center-of-mass frame.  This is sufficiently general as all Lorentz
violation is isolated in the wedge products which we never need to
explicitly evaluate.
We now turn to the individual contributions.

\begin{itemize}
\item { The four-point graph gives the following ${\cal O}\left( s^2 /m_{W}^4,
s/m_{W}^2 \right)$ contributions to the amplitude:
\vspace*{-1cm}
\begin{flushleft}
\begin{picture}(100,145)(10,10)
\Photon(31.72,66.56)(60,94.84){5}{3}
\put(20,99.84){$p_1$ $\searrow$}
\put(10,114.84){$W^{+}$}
\Photon(31.72,123.12)(60,94.84){5}{3}
\put(20,85.56){$p_2$ $\nearrow$}
\put(10,64.56){$W^{-}$}
\Photon(60,94.84)(88.28,123.12){5}{3}
\put(74,99.84){$\nearrow$ $p_3$}
\put(92,114.84){$W^{+}$}
\Photon(60,94.84)(88.28,66.56){5}{3}
\put(90,64.56){$W^{-}$}
\put(74,85.56){$\searrow$ $p_4$}
\put(110,94.84) {$= \left(ig^2 s^2 /16m_{W}^4 \right) \, \bigg\{ \, \left(1+
c_{\theta} \right)^2 \left[ {\rm cos}\left(p_1 \wedge p_2 -p_3 \wedge p_4 
\right) + {\rm cos}\left(p_1 \wedge p_3 -p_2 \wedge p_4 \right) \right]$}
\put(125,66.84) {$  - \left(1- c_{\theta} \right) ^2 {\rm cos}\left(p_1 \wedge
p_2 -p_3 \wedge p_4 \right)  -4 \, {\rm cos}\left(p_1 \wedge p_3 -p_2 \wedge 
p_4 \right) \, \bigg\}$}
\put(125,38.84) {$ + \left(ig^2 s /2m_{W}^2 \right) \, \bigg\{ \, \left(1-
c_{\theta} \right) {\rm cos}\left(p_1 \wedge p_3 -p_2 \wedge p_4 \right) $}
\put(125,10.84) {$ -2 \, c_{\theta} \, {\rm cos}\left(p_1 \wedge p_2 -p_3 
\wedge p_4 \right) \, \bigg\} \,\, , $} 
\end{picture}
\end{flushleft}
where we have introduced the abbreviation ${\rm cos}\left( \theta \right) =
c_{\theta}$, with $\theta$ being the scattering angle between the $p_1$ and 
$p_3$ three-momenta.  }
\item {The amplitude for the $s$-channel exchange for a generic gauge 
boson $V$ is given by
\vspace*{-1cm}
\begin{flushleft}
\begin{picture}(350,90)(10,10)
\Photon(63.28,48.28)(103.28,48.28){5}{3}
\put(78.28,58){$V$}
\Photon(103.28,48.28)(131.56,76.56){5}{3}
\put(117.28,53.28){$\nearrow$ $p_3$}
\put(138.28,68.28){$W^{+}$}
\Photon(103.28,48.28)(131.56,20){5}{3}
\put(138.28,18){$W^{-}$}
\put(117.28,39){$\searrow$ $p_4$}
\Photon(35,76.56)(63.28,48.28){5}{3}
\put(10,70){$W^{+}$}
\put(20,55){$p_1$ $\searrow$}
\Photon(35,20)(63.48,48.28){5}{3}
\put(10,18){$W^{-}$}
\put(20,38){$p_2$ $\nearrow$}
\put(165,48.28){$=\left(is^2 /4 m_{W}^4 \right)A^{21}_{V}A^{34}_{V} \, 
c_{\theta}  +\left( is m_{V}^2 /4 m_{W}^4 \right)A^{21}_{V}A^{34}_{V} \, 
c_{\theta} \,\, ,$} 
\end{picture}
\end{flushleft}
where the quantities $A_V^{ij}$ are obtained from the coefficients of
the gauge 3-point couplings in Fig. \ref{ncsmfeyn}, with the $ij$
denoting the momentum ordering in the wedge product.
For example, $A^{21}_{G^0} = g \, s_{23} \, {\rm sin}\left(p_2 \wedge
p_1 \right)$ and $A^{21}_{
\gamma} = -igs_{W} {\rm cos}\left(p_2 \wedge p_1 \right) -gc_{23}s_{11}c_{W}
{\rm sin}\left(p_2 \wedge p_1 \right)$. For the full amplitude we must sum 
over all possible intermediate states $V=\gamma, Z, G^0$ and $W^0$.}
\item {The $t$-channel
exchange of a generic gauge boson $V$ yields the following expression:
%
\begin{flushleft}
\begin{picture}(350,95)(10,10)
\Photon(30,80)(120,80){5}{6}
\Photon(75,80)(75,10){5}{4}
\Photon(30,10)(120,10){5}{6}
\put(10,78){$W^{+}$}
\put(35,66){$p_1$  $\rightarrow$}
\put(10,7){$W^{-}$}
\put(35,20){$p_2$  $\rightarrow$}
\put(128,78){$W^{+}$}
\put(95,66){$p_3$ $\rightarrow$}
\put(128,7){$W^{-}$}
\put(95,20){$p_4$  $\rightarrow$}
\put(85,45){$V$}
\put(160,45){$= -\left(is^2/16m_{W}^4 \right)A^{13}_{V}A^{42}_{V} \, \left(1-
c_{\theta} \right) \left(3+c_{\theta} \right) $}
\put(173,20){$+ \left(is/8m_{W}^2 \right)
A^{13}_{V}A^{42}_{V} \, \bigg\{ \, -16 c_{\theta} +\left( m_{V}^2 /m_{W}^2 
\right) \left(3+c_{\theta} \right) \, \bigg\} \,\, ,$}
\end{picture}
\end{flushleft}
where the $A_V^{ij}$ are as defined above. 
Again, to obtain the full amplitude we 
must sum over $V=\gamma, Z, G^0$ and $W^0$.}
\item {The $s$-channel SM Higgs exchange yields
\vspace*{-1cm}
\begin{flushleft}
\begin{picture}(150,90)(10,10)
\Line(63.28,48.28)(103.28,48.28)
\put(78.28,58){$h$}
\Photon(103.28,48.28)(131.56,76.56){5}{3}
\put(117.28,53.28){$\nearrow$ $p_3$}
\put(138.28,68.28){$W^{+}$}
\Photon(103.28,48.28)(131.56,20){5}{3}
\put(138.28,18){$W^{-}$}
\put(117.28,39){$\searrow$ $p_4$}
\Photon(35,76.56)(63.28,48.28){5}{3}
\put(10,70){$W^{+}$}
\put(20,55){$p_1$ $\searrow$}
\Photon(35,20)(63.48,48.28){5}{3}
\put(10,18){$W^{-}$}
\put(20,38){$p_2$ $\nearrow$}
\put(165,48.28){$= -\left(i g^2 s/4m_{W}^2 \right) {\rm exp}\left(i p_1 \wedge
p_2 -ip_3 \wedge p_4 \right) \,\, , $}
\end{picture}
\end{flushleft}}
\item {similarly the $t$-channel Higgs exchange gives the corresponding 
result 
%
\begin{flushleft}
\begin{picture}(300,95)(10,10)
\Photon(30,80)(120,80){5}{6}
\Line(75,80)(75,10)
\Photon(30,10)(120,10){5}{6}
\put(10,78){$W^{+}$}
\put(35,66){$p_1$  $\rightarrow$}
\put(10,7){$W^{-}$}
\put(35,20){$p_2$  $\rightarrow$}
\put(128,78){$W^{+}$}
\put(95,66){$p_3$ $\rightarrow$}
\put(128,7){$W^{-}$}
\put(95,20){$p_4$  $\rightarrow$}
\put(85,45){$h$}
\put(160,45){$= \left(i g^2 s/8 m_{W}^2 \right) \left(1-c_{\theta} \right)
{\rm exp}\left(-i p_1 \wedge p_3 +ip_2 \wedge p_4 \right) \,\, . $}
\end{picture}
\end{flushleft}}
\end{itemize}
Note that there are no Higgsac contributions as
they do not couple to the SM gauge fields.

Summing these expressions, using the relations
\begin{equation}
p_1 \wedge p_3 -p_2 \wedge p_4 = p_1 \wedge p_2 -p_3 \wedge p_4 \,\, ,
\label{ncsmid}
\end{equation}
which follows from momentum conservation and the 
antisymmetry of the wedge product, and $c_{23}s_{11}=\tan\theta_W\equiv t_{W}$, 
and setting $m_{W}^2 = m_{Z}^2 c_{W}^2$, we find that the ${\cal O}\left(s^2 /
m_{W}^4 \right)$ do indeed cancel as expected. 
The remaining ${\cal O}\left(s /m_{W}^2 \right)$ contributions only partially 
cancel, however, and the remainder can be combined to yield
\begin{eqnarray}
i{\cal M} &=& \frac{ig^2}{8} \frac{s}{m_{W}^2} \, \bigg\{ \, 6 \, {\rm cos}
\left(p_1 \wedge p_3 -p_2 \wedge p_4 \right) -3\, 
{\rm exp}\left(-ip_1 \wedge p_3 +ip_2 \wedge p_4 \right) 
\\ & & 
-3 \, \bigg[ \,  {\rm cos}\left(p_1 \wedge p_3 \right) \, {\rm cos}\left(
p_2 \wedge p_4 \right)  +t_{W}^4 \, {\rm sin}\left(p_1 \wedge p_3 \right) \, 
{\rm sin}\left(p_2 \wedge p_4 \right) \nonumber \\ & & 
+i \, t_{W}^2 \, {\rm sin}\left(p_1 \wedge p_3 - p_2 \wedge p_4 \right) \, 
\bigg]
-3\bigg[ \frac{m_{G^0}^2}{m_{W}^2} s_{23}^2 
+ \frac{m_{W^0}^2}{m_{W}^2} c_{23}^2 c_{11}^2\bigg] \, 
{\rm sin}\left(p_1 \wedge 
p_3 \right) \, {\rm sin}\left(p_2 \wedge p_4 \right) 
\nonumber \\ & & +c_{\theta} \, \bigg[ \,
4 \, {\rm cos}\left( p_1 \wedge p_2 -p_3 \wedge p_4 \right) 
-{\rm exp}\left(-ip_1 \wedge p_2 +i p_3 \wedge p_4 \right) \nonumber \\ 
& & - \, \bigg( \,
{\rm cos}\left(p_1 \wedge p_3 \right) \, {\rm cos}\left(p_2 \wedge p_4 \right)
+2 \, {\rm cos}\left(p_1 \wedge p_2 \right) \, {\rm cos}
\left(p_3 \wedge p_4 \right) +t_{W}^4 \, {\rm sin}\left(p_1 \wedge p_3 \right)
\, {\rm sin}\left(p_2 \wedge p_4 \right)  \nonumber \\ & &
+2 \, t_{W}^4 \, {\rm sin}\left(p_1 \wedge p_2 \right) \, {\rm sin}\left(p_3 
\wedge p_4 \right) -i \, t_{W}^2 \, {\rm sin}\left( p_1 \wedge p_3 -p_2 
\wedge p_4 \right) \, \bigg) \nonumber \\ & &
-\bigg[ \frac{m_{G^{0}}^2}{m_{W}^2} s_{23}^2 +
\frac{m_{W^{0}}^2}{m_{W}^2} c_{23}^2 c_{11}^2\bigg] \left\{
{\rm sin}\left(p_1 \wedge p_3 \right) \, {\rm sin}\left(p_2 \wedge p_4 \right)
+2 \, {\rm sin}\left(p_1 \wedge p_2 \right) \, {\rm sin}\left(p_3 \wedge p_4 
\right) \right\} \, \bigg] \, \bigg\} \nonumber \,\, .
\label{ncsmw4amp}
\end{eqnarray}
One can trivially check that in the commutative SM limit this expression 
indeed vanishes.  Remember that in the SM, the mass relationship between 
the $W^\pm$ and $Z$ 
fields as well as the Higgs exchanges ensures the cancellations. 
Although the above expression is rather complicated, it is clear 
that it cannot be made to 
vanish in the general NC case.  This is most 
easily seen by the presence of the large and most dangerous 
terms proportional to $m_{G^{0}}^2$ and $m_{W^{0}}^2$; recall 
that these masses are arbitrary functions of the two
Higgsac vevs. Here there are no corresponding relationships 
between the new gauge boson masses themselves or with those of the $W^\pm$ 
and $Z$ and there are no Higgsac contributions with couplings proportional to 
$m_{G^{0}}$ and/or $m_{W^{0}}$ to help the cancellations. As we see from the 
above calculation, 
the $m_{G^{0}}^2$ and $m_{W^{0}}^2$ terms arise from the $s$- and 
$t$-channel vector boson exchanges when the propagators are expanded to 
leading- and next-to-leading order in $s$. This means that such terms are 
unavoidable in this scenario and unitarity must fail.  

It is instructive to ask at what scale tree-level unitarity is violated in this 
model.  Clearly, this is dependent on the value of the center-of-mass energy,
since the theory is only unitary for all values of $\sqrt s$ when
$\Lambda_{NC}\to\infty$.
To be specific, let us examine the $W^+_LW^-_L$ scattering process 
in the center-of-mass frame and assume for simplicity that the non-commutativity
is of the space-space type. It is 
clear as noted above that the most dangerous terms are proportional to the 
squares of the new gauge boson masses. Keeping only these leading terms we find 
the constraint
\begin{equation}
\Lambda_B>0.83 [(\sqrt s)^6~\sin^2 \gamma~(m_{W^0}^2+0.19m_{G^0}^2)]^{1/4}\,,
\end{equation}
where all quantities are in TeV. Here $\gamma$ is the angle between the 
momentum $p_1$ and the unit vector $\hat c_B$. 
These bounds can be quite severe; taking
$\sin \gamma=1$ and $m_{W^0}=m_{G^0}$, one finds that 
$\Lambda_B >0.86(9.6,27.3)\sqrt m_{G^0}$ TeV for $\sqrt s=1(5,10)$ TeV,
which is a strong constraint for new gauge boson 
masses in the few TeV range.  Similar strong bounds are to be expected in the 
cases of space-time or mixed non-commutativity.

In order to elucidate the origin of the apparent problems that we have just 
encountered, it would be beneficial to examine a parallel set of processes in 
somewhat more simplified NC gauge theories that 
are more tractable.  By using a
set of test models we hope to cleanly isolate the basic causes of the 
unitarity failure in the NCSM and probe the issue in a more general way. 
We hope to answer the question of 
just when are spontaneously broken NC gauge theories non-unitary.

\section{Test Models}

In this section we examine gauge boson scattering in a variety of 
test cases of spontaneously broken NC theories.  We start with the
simplest possibility, that of spontaneously broken NC $U(1)$, and then
work our way towards more realistic models
adding one layer of complications at a time.  In this manner, we hope to
isolate the source of unitarity violation discovered above and determine
its consequences for future NC model building.  Since the 
unitarity failure encountered above occurred in the pure gauge/Higgs sector,
we need not worry about the fermion content of any of the test models that 
we examine below and thus need only consider the process of gauge boson 
scattering.

\subsection{Spontaneously Broken NC $U(1)$}

We first examine the simplest example of a spontaneously broken NC theory, that
of NC $U(1)$ which is broken by a complex scalar field transforming as a 
fundamental under the gauge group.  A one-loop analysis of this model was 
first performed in~\cite{renorm2}, to which the reader is referred 
for a more
detailed discussion.  The physical spectrum after symmetry breaking consists 
of a massive real scalar Higgs field $h$ and a massive vector field $Z$.  This 
theory was shown by explicit calculation to be one loop 
renormalizable\cite{renorm2}, and we therefore expect
it to be unitary at tree-level unlike the NCSM whose one-loop 
properties are unknown.  To test whether or not tree-level unitarity is 
satisfied we calculate the high energy limit of the scattering process
$Z_L Z_L \rightarrow Z_L Z_L$.  As in the NCSM case above,
we need to check if 
the potential unitarity violating terms of
${\cal O} \left( s^2 /m_{Z}^4, s/m_{Z}^2 \right)$ vanish after summing over 
all the diagrams. 

\begin{figure}
\begin{picture}(440,200)(10,10)
\Photon(20,190)(50,160){5}{3}
\Photon(20,130)(50,160){5}{3}
\Photon(50,160)(80,190){5}{3}
\Photon(50,160)(80,130){5}{3}
\put(10,172){$Z_1$}
\put(10,145){$Z_2$}
\put(75,172){$Z_3$}
\put(75,145){$Z_4$}

\Photon(120,185)(145,160){5}{3}
\Photon(120,130)(145,160){5}{3}
\Photon(145,160)(180,160){5}{3}
\Photon(180,160)(205,185){5}{3}
\Photon(180,160)(205,130){5}{3}
\put(110,170){$Z_1$}
\put(110,142){$Z_2$}
\put(205,170){$Z_3$}
\put(205,142){$Z_4$}

\Photon(245,185)(315,185){5}{6}
\Photon(280,185)(280,130){5}{4}
\Photon(245,130)(315,130){5}{6}
\put(245,170){$Z_1$}
\put(245,140){$Z_2$}
\put(305,170){$Z_3$}
\put(305,140){$Z_4$}

\Photon(355,185)(425,185){5}{6}
\Photon(390,185)(390,130){5}{4}
\Photon(355,130)(425,130){5}{6}
\put(355,170){$Z_1$}
\put(355,140){$Z_2$}
\put(415,170){$Z_4$}
\put(415,140){$Z_3$}

\Photon(70,75)(95,50){5}{3}
\Photon(70,20)(95,50){5}{3}
\Line(95,50)(130,50)
\Photon(130,50)(155,75){5}{3}
\Photon(130,50)(155,20){5}{3}
\put(60,60){$Z_1$}
\put(60,32){$Z_2$}
\put(155,60){$Z_3$}
\put(155,32){$Z_4$}

\Photon(195,75)(265,75){5}{6}
\Line(230,75)(230,20)
\Photon(195,20)(265,20){5}{6}
\put(195,60){$Z_1$}
\put(195,30){$Z_2$}
\put(255,60){$Z_3$}
\put(255,30){$Z_4$}

\Photon(305,75)(375,75){5}{6}
\Line(340,75)(340,20)
\Photon(305,20)(375,20){5}{6}
\put(305,60){$Z_1$}
\put(305,30){$Z_2$}
\put(365,60){$Z_4$}
\put(365,30){$Z_3$}
\end{picture}
\caption{Diagrams contributing to the scattering process $Z_L Z_L 
\rightarrow Z_L Z_L$ in NC $U(1)$; the subscripts denote the momentum carried
by the field.}
\label{u1z4diags}
\end{figure}
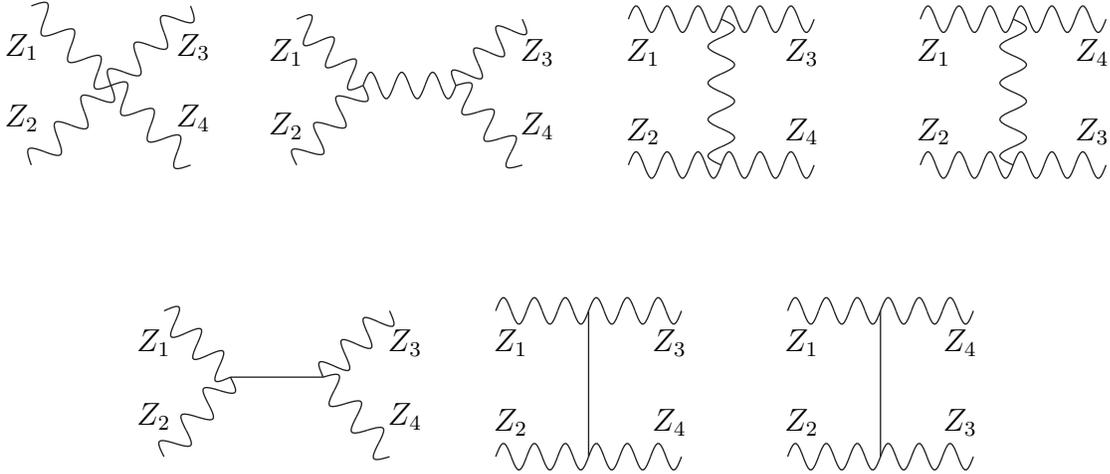

The Feynman rules
relevant to this calculation are given in Fig.~\ref{u1feyn}, and the
diagrams contributing to $Z_L Z_L \rightarrow Z_L Z_L$ are presented in
Fig.~\ref{u1z4diags}.  Summing these seven diagrams following the same 
procedure as in the case of the NCSM and keeping only the
${\cal O} \left( s^2 /m_{Z}^4, s/m_{Z}^2 \right)$ terms, yields the amplitude
\begin{eqnarray}
i{\cal M} &=& \frac{i g^2}{2} \frac{s}{m_{Z}^2}
\bigg\{ {\rm cos}(p_1 \wedge p_3 -p_2 \wedge p_4) + {\rm cos}(p_1 \wedge p_4 
-p_2 \wedge p_3) -2\, c_{(12)}c_{(34)} \nonumber \\ & & + c_{\theta} \bigg[ 
{\rm cos}(p_1 \wedge p_4
+ p_2 \wedge p_3) - {\rm cos}(p_1 \wedge p_3 + p_2 \wedge p_4) -10 \, \bigg(\, 
s_{(12)}s_{(34)} \nonumber \\ & & + s_{(14)}s_{(23)} -s_{(13)}s_{(24)} 
\bigg) \bigg] \bigg\} \,\, ,
\label{u1z4amp}
\end{eqnarray}
where $c_\theta$ is defined as in the previous section.
We have introduced the shorthand notation
${\rm sin}\left( p_i \wedge p_j \right) =s_{(ij)}$, ${\rm cos}\left( 
p_i \wedge p_j \right) =c_{(ij)}$; it is important that these should not be 
confused with the abbreviations used for the mixing angles in the NCSM.  
We see that the ${\cal O}(s^2 /m_{Z}^4)$ terms cancel trivially as 
expected and as in the NCSM case above.  
Use of the relations
\begin{eqnarray}
p_1 \wedge p_2 - p_3 \wedge p_4 &=& p_1 \wedge p_3 -p_2 \wedge p_4 \nonumber
\\
p_1 \wedge p_2 + p_3 \wedge p_4 &=& p_1 \wedge p_4 -p_2 \wedge p_3 \nonumber
\\
p_1 \wedge p_3 + p_2 \wedge p_4 &=& -p_1 \wedge p_4 -p_2 \wedge p_3 \,\, ,
\label{ids1}
\end{eqnarray}
which follow directly from momentum conservation and the antisymmetry of the
wedge product, and the consequent identity
\begin{equation}
s_{(12)}s_{(34)}=s_{(13)}s_{(24)}-s_{(14)}s_{(23)}\,,
\label{ids2}
\end{equation} 
then leads to a cancellation of the ${\cal O}(s /m_{Z}^2)$ terms. Unitarity is
thus preserved at high energies as expected for this scenario.  For the
sake of completeness, we also examined the process $hZ_L 
\rightarrow hZ_L$, and found that the leading ${\cal O}(s /m_{Z}^2)$ 
terms similarly cancel.  

This calculation demonstrates that it is possible to 
construct a spontaneously broken NC theory that preserves tree-level 
unitarity in 
processes involving the scattering of longitudinal gauge bosons. 
\begin{figure}
\begin{picture}(200,280)(10,10)
\put(10,56){$h$}
\Line(10,50)(50,50)
\Photon(50,50)(78.28,78.28){5}{3}
\put(64,55){$\nearrow$ $p_1$}
\put(82,74){$Z_{\mu}$}
\Photon(50,50)(78.28,21.72){5}{3}
\put(80,20){$Z_{\nu}$}
\put(64,41){$\searrow$ $p_2$}
\put(100,50){$=2ig \, m_Z {\rm cos}(p_1 \wedge p_2) g_{\mu \nu}$}

\Photon(10,148.28)(50,148.28){5}{3}
\put(10,158){$Z_{\mu}$}
\put(10,135){$\leftarrow$  $p_1$}
\Photon(50,148.28)(78.28,176.56){5}{3}
\put(64,153.28){$\nearrow$ $p_2$}
\put(82,168.28){$Z_{\nu}$}
\Photon(50,148.28)(78.28,120){5}{3}
\put(80,118){$Z_{\rho}$}
\put(64,139){$\searrow$ $p_3$}
\put(100,128.28){$= -2g \, {\rm sin}(p_1 \wedge p_2) \left[ (p_1 -p_2)_{\rho} 
g_{\mu\nu} +(p_2 -p_3)_{\mu}g_{\nu \rho} +(p_3 -p_1)_{\nu}g_{\mu\rho}\right]$}

\Photon(21.72,274.84)(50,246.56){5}{3}
\put(10,251.56){$p_1$ $\nwarrow$}
\put(10,266.56){$Z_{\mu}$}
\Photon(21.72,218.28)(50,246.56){5}{3}
\put(10,237.28){$p_2$ $\swarrow$}
\put(10,216.28){$Z_{\nu}$}
\Photon(50,246.56)(78.28,274.84){5}{3}
\put(64,251.56){$\nearrow$ $p_3$}
\put(82,266.56){$Z_{\sigma}$}
\Photon(50,246.56)(78.28,218.28){5}{3}
\put(80,216.28){$Z_{\rho}$}
\put(64,237.28){$\searrow$ $p_4$}
\put(100,246.56){$= -4ig^2 \, \bigg\{ {\rm sin}(p_1 \wedge p_2)\, {\rm sin}
(p_3 \wedge p_4) \left(g_{\mu \sigma}g_{\nu \rho}-g_{\mu \rho}g_{\nu \sigma}
\right) + $}
\put(153,225){${\rm sin}(p_3 \wedge p_1)\, {\rm sin}
(p_2 \wedge p_4) \left(g_{\mu \rho}g_{\nu \sigma}-g_{\mu \nu}g_{\rho \sigma}
\right) + $}
\put(153,204){${\rm sin}(p_1 \wedge p_4)\, {\rm sin}
(p_2 \wedge p_3) \left(g_{\mu \nu}g_{\rho \sigma}-g_{\mu \sigma}g_{\nu \rho}
\right) \bigg\} $}
\end{picture}
\caption{NC $U(1)$ Feynman rules relevant for $Z_L Z_L \rightarrow Z_L Z_L$.}
\label{u1feyn}
\end{figure}
\vspace*{-1.0cm}

\bigskip

\subsection{$U(1) \times U(1)$}

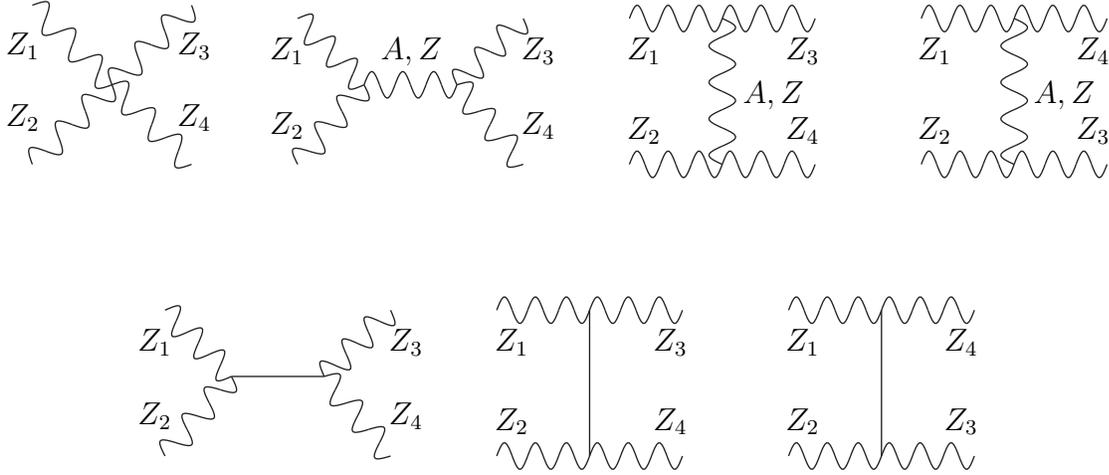
\begin{figure}
\begin{picture}(440,200)(10,10)
\Photon(20,190)(50,160){5}{3}
\Photon(20,130)(50,160){5}{3}
\Photon(50,160)(80,190){5}{3}
\Photon(50,160)(80,130){5}{3}
\put(10,172){$Z_1$}
\put(10,145){$Z_2$}
\put(75,172){$Z_3$}
\put(75,145){$Z_4$}

\Photon(120,185)(145,160){5}{3}
\Photon(120,130)(145,160){5}{3}
\Photon(145,160)(180,160){5}{3}
\Photon(180,160)(205,185){5}{3}
\Photon(180,160)(205,130){5}{3}
\put(110,170){$Z_1$}
\put(110,142){$Z_2$}
\put(205,170){$Z_3$}
\put(205,142){$Z_4$}
\put(152,170){$A,Z$}

\Photon(245,185)(315,185){5}{6}
\Photon(280,185)(280,130){5}{4}
\Photon(245,130)(315,130){5}{6}
\put(245,170){$Z_1$}
\put(245,140){$Z_2$}
\put(305,170){$Z_3$}
\put(305,140){$Z_4$}
\put(289,153.5){$A,Z$}

\Photon(355,185)(425,185){5}{6}
\Photon(390,185)(390,130){5}{4}
\Photon(355,130)(425,130){5}{6}
\put(355,170){$Z_1$}
\put(355,140){$Z_2$}
\put(415,170){$Z_4$}
\put(415,140){$Z_3$}
\put(399,153.5){$A,Z$}

\Photon(70,75)(95,50){5}{3}
\Photon(70,20)(95,50){5}{3}
\Line(95,50)(130,50)
\Photon(130,50)(155,75){5}{3}
\Photon(130,50)(155,20){5}{3}
\put(60,60){$Z_1$}
\put(60,32){$Z_2$}
\put(155,60){$Z_3$}
\put(155,32){$Z_4$}

\Photon(195,75)(265,75){5}{6}
\Line(230,75)(230,20)
\Photon(195,20)(265,20){5}{6}
\put(195,60){$Z_1$}
\put(195,30){$Z_2$}
\put(255,60){$Z_3$}
\put(255,30){$Z_4$}

\Photon(305,75)(375,75){5}{6}
\Line(340,75)(340,20)
\Photon(305,20)(375,20){5}{6}
\put(305,60){$Z_1$}
\put(305,30){$Z_2$}
\put(365,60){$Z_4$}
\put(365,30){$Z_3$}
\end{picture}
\caption{Diagrams contributing to the scattering process $Z_L Z_L 
\rightarrow Z_L Z_L$ in NC $U(1) \times U(1)$; the subscripts denote the 
momentum carried by the field.}
\label{u1u1z4diags}
\end{figure}

Perhaps $U(1)$ is too simple of an example to reveal the potential 
unitarity failure of spontaneously broken NC theories.
We next extend our investigation to
a NC $U(1) \times U(1)$ theory; this example contains a direct
product gauge group and mixing between the gauge boson of each group, which
introduces additional complexities beyond the model of Sec. 3.1.  The
Lagrangian density defining this theory is
\begin{equation}
{\cal L}= -\frac{1}{4} \, B_{\mu\nu} * B^{\mu\nu} - \frac{1}{4} \, 
C_{\mu\nu} * C^{\mu\nu} + \left(D_{\mu}\phi \right)^{\dagger} *
D^{\mu} \phi -V \left(\phi \right)\,\, ,
\label{u1u1lag}
\end{equation}
where $B_{\mu\nu}=\partial_{\mu}B_{\nu}-\partial_{\nu}B_{\mu} +ig \, [B_{\mu},
B_{\nu}]_{MB}$ and similarly for $C_{\mu\nu}$ with $g\to g'$, 
$D_{\mu}\phi=\partial_{\mu}\phi
+ig\,B_{\mu} * \phi-ig^{'}\phi * C_{\mu}$, and the potential
$V \left(\phi \right)$
is chosen so that its minimum is at $\phi_0 = \nu$.  Expanding $\phi$
around $\nu$, and introducing the suggestive notation
\begin{equation}
s_{W}=\frac{g}{\sqrt{g^2 +g^{'2}}}\, , \,\, c_{W}=\frac{g^{'}}{\sqrt{g^2 +
g^{'2}}}\, , \,\, e=\sqrt{g^2 +g^{'2}} \,\, ,
\label{u1u1not}
\end{equation}
we find a physical spectrum of fields consisting of a real 
massive scalar Higgs $h$ and the following gauge bosons:
\begin{eqnarray}
A &=& c_{W}B + s_{W}C \, , \,\,\, m_{A}^2 =0 \,\, ; \nonumber \\
Z &=& s_{W}B - c_{W}C \, , \,\,\, m_{Z}^2 =2e^2 \nu^2 \,\, .
\label{u1u1spec}
\end{eqnarray}
We again consider the scattering process $Z_L Z_L \rightarrow Z_L Z_L$, and
determine whether the unitarity violating ${\cal O} \left( s^2 /m_{Z}^4, 
s/m_{Z}^2 \right)$ terms vanish.  We present the necessary Feynman rules 
for this scenario in Fig.~\ref{u1u1feyn}, and the contributing diagrams 
to this process in Fig.~\ref{u1u1z4diags}. 
Note that exchanges of multiple gauge bosons in the $s$- and $t$-channels are 
now present since an $AZZ$ 3-point coupling is induced through mixing. 
In deriving these Feynman rules we have used the trigonometric relations
\begin{equation}
s_{W}^4 -c_{W}^4 = s_{W}^2 -c_{W}^2 \,\, , \,\,\, s_{W}^6 +c_{W}^6 =1-3s_{W}^2 
c_{W}^2 \,\, .
\label{u1u1ids}
\end{equation}
Summing the diagrams in Fig.~\ref{u1u1z4diags}, and keeping only ${\cal O} 
\left( s^2 /m_{Z}^4, s/m_{Z}^2 \right)$ contributions, we arrive at the 
amplitude
\begin{eqnarray}
i{\cal M} &=& -ie^2 \frac{s^2}{m_{Z}^4} \bigg\{ \left[\left(-1+3s_{W}^2 
c_{W}^2 \right) +s_{W}^4 -s_{W}^2 c_{W}^2 +c_{W}^4 \right] \bigg( c_{\theta}
s_{(12)}s_{(34)} \nonumber \\ 
& & +\frac{1}{4}(-3+2c_{\theta} +c_{\theta}^2 \,
)\, s_{(13)}s_{(24)} +\frac{1}{4}(-3-2c_{\theta}+c_{\theta}^2 \,)\, 
s_{(14)}s_{(23)}
\bigg)\bigg\} \nonumber \\ 
& & + \frac{ie^2}{2} \frac{s}{m_{Z}^2} \bigg\{4 \left(1-3s_{W}^2 
c_{W}^2 \right)  \left(s_{(13)}s_{(24)}+s_{(14)}s_{(23)} \right) \nonumber \\ 
& & -3 \left(1-4s_{W}^2 c_{W}^2 
\right)\left(s_{(13)}s_{(24)}+s_{(14)}s_{(23)} \right)
-2c_{(12)}c_{(34)} +c_{(13)}c_{(24)}+c_{(14)}c_{(23)} \nonumber \\ & &
+c_{\theta} \bigg[\left( 11-32s_{W}^2 c_{W}^2 \right) \left(s_{(13)}s_{(24)}
-s_{(14)}s_{(23)} \right) -2 \left(1-4s_{W}^2 c_{W}^2 \right)s_{(12)}s_{(34)}
\nonumber \\ & & -8\left( 1-3s_{W}^2 c_{W}^2 \right)s_{(12)}s_{(34)} 
+c_{(14)}c_{(23)}-c_{(13)}c_{(24)} \bigg] \bigg\} \,\, .
\label{u1u1z4amp}
\end{eqnarray}
The ${\cal O}\left( s^2 /m_{Z}^4 \right)$ terms have vanished using the 
relation $s_{W}^4 +c_{W}^4 = 1-2s_{W}^2 c_{W}^2$ and the ${\cal O}\left( s 
/m_{Z}^2 \right)$ contributions also cancel by 
using the relations in Eqs.~\ref{ids1} and~\ref{ids2}. Thus the presence of 
product groups and multiple gauge bosons is {\it not} the crucial feature 
causing the lack of tree-level unitarity in the NCSM. Again, for completeness,
we have also studied the process $AZ_L \rightarrow AZ_L$, 
and found that the leading ${\cal O}\left( s /m_{Z}^2 \right)$ terms cancel. 

We therefore conclude that the 
additional complexities of direct product gauge groups and gauge boson mixing
do not destroy our ability to maintain unitarity in NC processes containing
longitudinal vector particles.
\begin{figure}
\begin{picture}(200,385)(10,10)
\put(10,56){$h$}
\Line(10,50)(50,50)
\Photon(50,50)(78.28,78.28){5}{3}
\put(64,55){$\nearrow$ $p_1$}
\put(82,74){$Z_{\mu}$}
\Photon(50,50)(78.28,21.72){5}{3}
\put(80,20){$Z_{\nu}$}
\put(64,41){$\searrow$ $p_2$}
\put(100,50){$=2ie \, m_Z {\rm cos}(p_1 \wedge p_2) g_{\mu \nu}$}

\Photon(10,148.28)(50,148.28){5}{3}
\put(10,158){$Z_{\mu}$}
\put(10,135){$\leftarrow$  $p_1$}
\Photon(50,148.28)(78.28,176.56){5}{3}
\put(64,153.28){$\nearrow$ $p_2$}
\put(82,168.28){$Z_{\nu}$}
\Photon(50,148.28)(78.28,120){5}{3}
\put(80,118){$Z_{\rho}$}
\put(64,139){$\searrow$ $p_3$}
\put(100,128.28){$= -2e \left(s_{W}^2 -c_{W}^2 \right) {\rm sin}(p_1 
\wedge p_2) \left[ (p_1 -p_2)_{\rho} g_{\mu\nu} +(p_2 -p_3)_{\mu}g_{\nu \rho} 
+(p_3 -p_1)_{\nu}g_{\mu\rho}\right]$}

\Photon(10,246.56)(50,246.56){5}{3}
\put(10,256.28){$A_{\mu}$}
\put(10,233.28){$\leftarrow$  $p_1$}
\Photon(50,246.56)(78.28,274.84){5}{3}
\put(64,251.56){$\nearrow$ $p_2$}
\put(82,266.56){$Z_{\nu}$}
\Photon(50,246.56)(78.28,218.28){5}{3}
\put(80,216.28){$Z_{\rho}$}
\put(64,236.28){$\searrow$ $p_3$}
\put(100,225.56){$= -2e \, s_{W}c_{W} \, {\rm sin}(p_1 
\wedge p_2) \left[ (p_1 -p_2)_{\rho} g_{\mu\nu} +(p_2 -p_3)_{\mu}g_{\nu \rho} 
+(p_3 -p_1)_{\nu}g_{\mu\rho}\right]$}

\Photon(21.72,374.84)(50,346.56){5}{3}
\put(10,351.56){$p_1$ $\nwarrow$}
\put(10,366.56){$Z_{\mu}$}
\Photon(21.72,318.28)(50,346.56){5}{3}
\put(10,337.28){$p_2$ $\swarrow$}
\put(10,316.28){$Z_{\nu}$}
\Photon(50,346.56)(78.28,374.84){5}{3}
\put(64,351.56){$\nearrow$ $p_3$}
\put(82,366.56){$Z_{\sigma}$}
\Photon(50,346.56)(78.28,318.28){5}{3}
\put(80,316.28){$Z_{\rho}$}
\put(64,337.28){$\searrow$ $p_4$}
\put(100,346.56){$= -4ie^2 \left(1-3s_{W}^2 c_{W}^2 \right) \bigg\{ {\rm sin}
(p_1 \wedge p_2)\, {\rm sin}
(p_3 \wedge p_4) \left(g_{\mu \sigma}g_{\nu \rho}-g_{\mu \rho}g_{\nu \sigma}
\right) + $}
\put(153,325){${\rm sin}(p_3 \wedge p_1)\, {\rm sin}
(p_2 \wedge p_4) \left(g_{\mu \rho}g_{\nu \sigma}-g_{\mu \nu}g_{\rho \sigma}
\right) + $}
\put(153,304){${\rm sin}(p_1 \wedge p_4)\, {\rm sin}
(p_2 \wedge p_3) \left(g_{\mu \nu}g_{\rho \sigma}-g_{\mu \sigma}g_{\nu \rho}
\right) \bigg\} $}
\end{picture}
\caption{NC $U(1) \times U(1)$ Feynman rules relevant for $Z_L Z_L 
\rightarrow Z_L Z_L$.}
\label{u1u1feyn}
\end{figure}
\vspace*{-1.0cm}

\bigskip
\subsection{$U(2)$: Fundamental $+$ Adjoint Breaking}

Perhaps unitarity was satisfied for the previous two test cases because
both $U(1)$ and $U(1) \times U(1)$ are abelian 
gauge groups.  To address this issue, 
we study NC $U(2)$ broken by both a fundamental and an
adjoint Higgs; this example furnishes the breaking of a non-abelian gauge 
group together with a more complicated example of mixing
than that found in the $U(1) \times U(1)$ case.  It contains 
a two step symmetry breaking
process akin to that used in the NCSM of~\cite{Chaichian:2001py} as well as a 
spectrum which closely resembles that in~\cite{Chaichian:2001py}.  The 
theory is defined by the Lagrangian density
\begin{equation}
{\cal L}= -\frac{1}{2}{\rm Tr} \left(B_{\mu\nu} * B^{\mu\nu} \right)
+\left( D_{\mu}\phi_F \right)^{\dagger} * D^{\mu} \phi_F -V\left(\phi_F 
\right) +{\rm Tr} \left(D_{\mu}\phi_A * D^{\mu}\phi_A \right) -V'\left(
\phi_A \right) \,\, ,
\label{u2falag}
\end{equation}
where $\phi_F$ is the fundamental Higgs, $\phi_A$ the matrix valued adjoint
Higgs, and $B_{\mu}$ the matrix valued $U(2)$ gauge field.  The field strength
tensor and covariant derivatives are 
\begin{eqnarray}
B_{\mu\nu} &=& \partial_{\mu}B_{\nu}-\partial_{\nu}B_{\mu}+g \left[B_{\mu},
B_{\nu} \right]_{MB} \,\, , \nonumber \\ 
D_{\mu} \phi_F &=& \partial_{\mu} \phi_F +ig \, B_{\mu} * \phi_F \,\, ,
\nonumber \\
D_{\mu} \phi_A &=& \partial_{\mu} \phi_A +g \left[ B_{\mu}, \phi_A 
\right]_{MB}\,\, .
\label{u2fadefs}
\end{eqnarray}
We choose the potentials $V\left(\phi_F \right), V'\left(\phi_A \right)$
so that their symmetry breaking minima occur at 
\begin{equation}
\phi_{F,0}= \nu \, \left( \begin{array}{c} 0 \\ 1 \end{array} \right)  \,\, , 
\,\,\, \phi_{A,0}=\frac{\nu^{'}}{\sqrt{2}} \, \frac{\sigma_{3}}{2} \,\, .
\label{u2famin}
\end{equation}
The gauge sector of this theory consists of the following particles:
\begin{eqnarray}
A &=& \frac{1}{\sqrt{2}} \left(B_0 +B_3 \right) \,\, , \,\,\, 
m^2 =0 \nonumber \\
Z &=& \frac{1}{\sqrt{2}} \left(B_0 -B_3 \right) \,\, , \,\,\, 
m_{Z}^2 =g^2 \nu^2 \nonumber \\
W^{\pm} &=& \frac{1}{\sqrt{2}} \left(B_1 \mp iB_2 \right) \,\, , \,\,\, 
m_{W}^2 =m_{F}^2 +m_{A}^2 \,\, , \nonumber \\
& & m_{F}^2 = \frac{g^2 \nu^2}{2} \,\, , \,\,\, m_{A}^2 =\frac{g^2 \nu^{'2}}
{2} \,\, .
\label{u2faspec}
\end{eqnarray}
Our calculation will involve the contributions from the 
two neutral scalars $\phi_0 , \phi_3 $ 
from $\phi_A =  \phi_{\mu} \, \sigma^{\mu}/2$, and the scalar $h$ from the 
expansion $\phi_F= \nu+h/\sqrt{2} +i \sigma/\sqrt{2}$.  We note in particular 
the relation $m_{Z}^2 = 2 m_{F}^2$, which will be used later in this section.

As above, we examine the process $W^{+}_{L}
W^{-}_{L} \rightarrow W^{+}_{L} W^{-}_{L}$.  The pertinent Feynman rules for
this calculation are presented in Fig.~\ref{u2fafeyn}, and the relevant
diagrams in Fig.~\ref{u2faw4diags}.  Summing these contributions, we
find that the leading terms again cancel leaving the 
following ${\cal O}\left(s/m_{W}^2 \right)$ contribution to the amplitude:
(Note that the ${\cal O}\left(s^2 /m_{W}^4 \right)$ terms cancel after 
straightforward manipulations and so have not been displayed for this case.) 
\begin{eqnarray}
i{\cal M} &=& \frac{ig^2}{16} \frac{s}{m_{W}^2} \, 
\bigg\{ 8\left(c_{(13)}c_{(24)}+s_{(13)}s_{(24)} \right) -3 
\frac{m_{Z}^2}{m_{W}^2} {\rm exp}\left(-ip_1 \wedge 
p_3 +i p_2 \wedge p_4\right) \nonumber \\ & & 
-4 \frac{m_{F}^2}{m_{W}^2} {\rm exp}\, ( \, ip_1 
\wedge p_2 
-i p_3 \wedge p_4 \, ) \, 
+2 \frac{m_{F}^2}{m_{W}^2} {\rm exp}\left(-ip_1 \wedge p_3 +i p_2 \wedge 
p_4\right) \nonumber \\ & & 
-16 \frac{m_{A}^2}{m_{W}^2} \left(c_{(12)}c_{(34)}+s_{(12)}s_{(34)} 
\right)
+8 \frac{m_{A}^2}{m_{W}^2}  \left(c_{(13)}c_{(24)}+s_{(13)}s_{(24)} 
\right)\nonumber \\ & &  +c_{\theta}
\, \bigg[ -16 \left( c_{(12)}c_{(34)}+s_{(12)}s_{(34)} \right) +24 
\left( c_{(13)}c_{(24)}
+s_{(13)}s_{(24)} \right) \nonumber \\ & &
-2 \frac{m_{Z}^2}{m_{W}^2} {\rm exp}\left( ip_1 \wedge p_2 -ip_3 \wedge p_4
\right) -\frac{m_{Z}^2}{m_{W}^2} {\rm exp}\left( -ip_1 \wedge p_3 +ip_2 \wedge 
p_4 \right) \nonumber \\ & &
-2 \frac{m_{F}^2}{m_{W}^2} {\rm exp}\left( -ip_1 \wedge p_3 +ip_2 \wedge 
p_4 \right) -8 \frac{m_{A}^2}{m_{W}^2} \left( c_{(13)}c_{(24)}+s_{(13)}s_{(24)} 
\right)\bigg] \bigg\}
\label{u2w4amp}
\end{eqnarray}
Although messy in appearance this potentially unitarity violating 
contribution can be shown to vanish through the use of the relations in 
Eq.~\ref{ids1} and the identities $m_{Z}^2 = 2m_{F}^2$, $m_{W}^2 =m_{F}^2 +
m_{A}^2$. Here we see that the existence of certain mass relationships can be 
crucial in obtaining unitarity. 
We have also verified the cancellation of these dangerous terms 
in the scattering processes $hZ_L \rightarrow hZ_L$,
$hW^{+}_L \rightarrow hW^{+}_L$, and $W^{+}_T W^{-}_L \rightarrow W^{+}_T 
W^{-}_L$.  

The additional features present in this simplified theory do not ruin the 
delicate cancellations required to maintain unitarity; the problem with the 
NCSM of~\cite{Chaichian:2001py} must lie elsewhere, even though this toy model 
has features similar to those of the NCSM. 
\begin{figure}
\begin{picture}(350,485)(10,10)
\put(10,56){$h$}
\Line(10,50)(50,50)
\Photon(50,50)(78.28,78.28){5}{3}
\put(64,55){$\nearrow$ $p_1$}
\put(82,74){$W^{+}_{\mu}$}
\Photon(50,50)(78.28,21.72){5}{3}
\put(80,20){$W^{-}_{\nu}$}
\put(64,41){$\searrow$ $p_2$}
\put(100,50){$=ig \, m_F e^{i\, p_1 \wedge p_2} g_{\mu \nu}$}

\put(230,56){$\phi_3$}
\Line(230,50)(270,50)
\Photon(270,50)(298.28,78.28){5}{3}
\put(284,55){$\nearrow$ $p_1$}
\put(302,74){$W^{+}_{\mu}$}
\Photon(270,50)(298.28,21.72){5}{3}
\put(300,20){$W^{-}_{\nu}$}
\put(284,41){$\searrow$ $p_2$}
\put(320,50){$=2ig \, m_A \, {\rm cos}\left( p_1 \wedge p_2\right) 
g_{\mu \nu}$}

\put(10,154.28){$\phi_0$}
\Line(10,148.28)(50,148.28)
\Photon(50,148.28)(78.28,176.56){5}{3}
\put(64,153.28){$\nearrow$ $p_1$}
\put(82,172.28){$W^{+}_{\mu}$}
\Photon(50,148.28)(78.28,120){5}{3}
\put(80,118.28){$W^{-}_{\nu}$}
\put(64,139.28){$\searrow$ $p_2$}
\put(100,148.28){$=2g \, m_A \, {\rm sin}\left( p_1 \wedge p_2\right) 
g_{\mu \nu}$}

\Photon(10,246.56)(50,246.56){5}{3}
\put(10,256.28){$Z_{\rho}$}
\put(10,233.28){$\leftarrow$  $p_3$}
\Photon(50,246.56)(78.28,274.84){5}{3}
\put(64,251.56){$\nearrow$ $p_1$}
\put(82,266.56){$W^{-}_{\mu}$}
\Photon(50,246.56)(78.28,218.28){5}{3}
\put(80,216.28){$W^{+}_{\nu}$}
\put(64,237.28){$\searrow$ $p_2$}
\put(100,246.56){$= -\frac{ig}{\sqrt{2}} e^{-i\, p_1 \wedge p_2}
 \left[ (p_1 -p_2)_{\rho} g_{\mu\nu} +(p_2 -p_3)_{\mu}g_{\nu \rho} 
+(p_3 -p_1)_{\nu}g_{\mu\rho}\right]$}

\Photon(10,344.84)(50,344.84){5}{3}
\put(10,354.56){$A_{\rho}$}
\put(10,331.56){$\leftarrow$  $p_3$}
\Photon(50,344.84)(78.28,373.12){5}{3}
\put(64,349.84){$\nearrow$ $p_1$}
\put(82,364.84){$W^{-}_{\mu}$}
\Photon(50,344.84)(78.28,316.56){5}{3}
\put(80,314.56){$W^{+}_{\nu}$}
\put(64,334.56){$\searrow$ $p_2$}
\put(100,344.84){$=  \frac{ig}{\sqrt{2}} e^{i\, p_1 \wedge p_2}
\left[ (p_1 -p_2)_{\rho} g_{\mu\nu} +(p_2 -p_3)_{\mu}g_{\nu \rho} 
+(p_3 -p_1)_{\nu}g_{\mu\rho}\right]$}

\Photon(31.72,473.12)(60,444.84){5}{3}
\put(20,449.84){$p_1$ $\nwarrow$}
\put(10,464.84){$W^{+}_{\mu}$}
\Photon(31.72,416.56)(60,444.84){5}{3}
\put(20,435.56){$p_3$ $\swarrow$}
\put(10,414.56){$W^{+}_{\nu}$}
\Photon(60,444.84)(88.28,473.12){5}{3}
\put(74,449.84){$\nearrow$ $p_4$}
\put(92,464.84){$W^{-}_{\sigma}$}
\Photon(60,444.84)(88.28,416.56){5}{3}
\put(90,414.56){$W^{-}_{\rho}$}
\put(74,435.56){$\searrow$ $p_2$}
\put(110,444.84){$= ig^2 \, \bigg\{ \left[{\rm cos}\left(p_1 \wedge p_2 +
p_3 \wedge p_4 \right) +{\rm cos}\left(p_1 \wedge p_4 +
p_2 \wedge p_3 \right) \right]g_{\mu\nu}g_{\rho\sigma}$}
\put(143,423.28){$-{\rm cos}\left( p_1 \wedge p_4 +p_2 \wedge p_3 \right) 
g_{\mu\rho}g_{\nu\sigma} $}
\put(143,401.72){$-{\rm cos}\left( p_1 \wedge p_2 +p_3 \wedge p_4 \right) 
g_{\mu\sigma}g_{\nu\rho} \, \bigg\} $}
\end{picture}
\caption{NC $U(2)$ Feynman rules relevant for $W^{+}_L W^{-}_L 
\rightarrow W^{+}_L W^{-}_L$.}
\label{u2fafeyn}
\end{figure}
\vspace*{-1.0cm}
\begin{figure}
\begin{picture}(440,200)(10,10)
\Photon(80,190)(110,160){5}{3}
\Photon(80,130)(110,160){5}{3}
\Photon(110,160)(140,190){5}{3}
\Photon(110,160)(140,130){5}{3}
\put(62,172){$W^{+}_1$}
\put(62,145){$W^{-}_2$}
\put(135,172){$W^{+}_3$}
\put(135,145){$W^{-}_4$}

\Photon(190,185)(215,160){5}{3}
\Photon(190,130)(215,160){5}{3}
\Photon(215,160)(250,160){5}{3}
\Photon(250,160)(275,185){5}{3}
\Photon(250,160)(275,130){5}{3}
\put(172,170){$W^{+}_1$}
\put(172,142){$W^{-}_2$}
\put(278,170){$W^{+}_3$}
\put(278,142){$W^{-}_4$}
\put(222,170){$A,Z$}

\Photon(325,185)(395,185){5}{6}
\Photon(360,185)(360,130){5}{4}
\Photon(325,130)(395,130){5}{6}
\put(325,170){$W^{+}_1$}
\put(325,140){$W^{-}_2$}
\put(385,170){$W^{+}_3$}
\put(385,140){$W^{-}_4$}
\put(369,155.5){$A,Z$}

\Photon(130,75)(155,50){5}{3}
\Photon(130,20)(155,50){5}{3}
\Line(155,50)(190,50)
\Photon(190,50)(215,75){5}{3}
\Photon(190,50)(215,20){5}{3}
\put(112,60){$W^{+}_1$}
\put(112,32){$W^{-}_2$}
\put(218,60){$W^{+}_3$}
\put(218,32){$W^{-}_4$}
\put(155,60){$h,\phi_0,\phi_3$}

\Photon(265,75)(335,75){5}{6}
\Line(300,75)(300,20)
\Photon(265,20)(335,20){5}{6}
\put(265,60){$W^{+}_1$}
\put(265,30){$W^{-}_2$}
\put(325,60){$W^{+}_3$}
\put(325,30){$W^{-}_4$}
\put(305,46.5){$h,\phi_0,\phi_3$}

\end{picture}
\caption{Diagrams contributing to the scattering process $W^{+}_L W^{-}_L 
\rightarrow W^{+}_L W^{-}_L$ in NC U(2); the subscripts denote the 
momentum carried by the field.}
\label{u2faw4diags}
\end{figure}

\bigskip
\subsection{$U(2)$: Fundamental Higgs $+$ Higgsac}

We now examine NC $U(2)$ broken by both a fundamental Higgs
and a Higgsac.  Although this model is identical to that studied in Sec. 3.3
upon replacement of the Higgsac with an adjoint Higgs, we will find that the 
dangerous ${\cal O}\left(s /m_{W}^2 \right)$ terms now remain in the 
amplitude for
$W^{+}_L W^{-}_L \rightarrow W^{+}_L W^{-}_L$.  This theory is also quite
similar to the NCSM of~\cite{Chaichian:2001py}, the only difference here being
that the Higgsac breaks a $U(1)$ subgroup of $U(2)$ rather than a $U(1)$ 
subgroup of $U(2) \times U(1)$.  The success of the model of Sec. 3.3 
therefore allows us to conclude that the problem with the theory proposed 
in~\cite{Chaichian:2001py} is the use of the Higgsacs to break the NC $U(1)$
subgroups of $U(3) \times U(2) \times U(1)$.

We begin with the Lagrangian density
\begin{equation}
{\cal L}= -\frac{1}{2} {\rm Tr} \left( B_{\mu\nu} * B^{\mu\nu} \right)
+\left(D_{\mu}\phi_F \right)^{\dagger} * D^{\mu} \phi_F -V \left( \phi_F
\right) +\left(D_{\mu}\phi_H \right)^{\dagger} * D^{\mu} \phi_H -
V' \left( \phi_H \right) \,\, .
\label{u2fhlag}
\end{equation}
The field strength tensor $B_{\mu\nu}$ and the covariant derivative of the
fundamental Higgs $\phi_F$ are the same as given in Eq.~\ref{u2fadefs} of
Sec. 3.3, and the covariant derivative of the Higgsac is given by
\begin{equation}
D_{\mu} \phi_H = \partial_{\mu} \phi_H +\frac{ig}{2} B^{0}_{\mu} *
\phi_H \,\, .
\label{u2fhcov}
\end{equation}
Our conventions for the Higgsac covariant derivative differ slightly from
those in~\cite{Chaichian:2001py} in that we have not included a factor of
two in the interaction term arising from the trace of the SU(2) identity
matrix.  This is consistent with our normalization of the triple gauge
couplings.  We choose the same minimum for $\phi_F$ as in Eq.~\ref{u2famin}, 
and the Higgsac minimum at $\phi_{H,0}=\nu^{'}$.  The spectrum of Sec. 3.3 
consisted of a massless boson $A$ and three massive bosons $Z,W^{\pm}$; the 
effect of the Higgsac here is to induce mixing between $A$ and $Z$, 
leading to the following gauge sector spectrum:
\begin{eqnarray}
W^{\pm} &=& \frac{1}{\sqrt{2}}\left(B_1 \mp iB_2 \right) \,\, , \,\,\,
m_{W}^2 = \frac{g^2 \nu^2}{2} \,\, , \nonumber \\
Z^{+} &=& c_{\alpha}Z +s_{\alpha}A \,\, , \,\,\, m_{+}^2 = m_{W}^2 \left(
1+ {\rm sec}\left(\beta \right) +{\rm tan}\left(\beta \right) 
\right) \,\, , \nonumber \\
Z^{-} &=& c_{\alpha}A -s_{\alpha}Z \,\, , \,\,\, m_{-}^2 =  m_{W}^2 \left(
1- {\rm sec}\left(\beta \right) +{\rm tan}\left(\beta \right) \right)\,\, .
\label{u2fhspec}
\end{eqnarray}
Here $s_{\alpha},c_{\alpha}$ are shorthand for the mixing angles ${\rm sin} 
\left( \alpha \right), {\rm cos} \left( \alpha \right)$, and
${\rm tan}\left( \beta \right) $ is the ratio of the Higgsac and fundamental
Higgs vevs; the explicit expressions are given by 
\begin{eqnarray}
{\rm tan}\left( \beta \right) &=& \frac{\nu^{'2}}{2\nu^2} \,\, , \nonumber \\
\alpha &=& \frac{\beta}{2} \,\, .
\label{u2fhdefs}
\end{eqnarray}
Upon taking $\beta \rightarrow 0$ we recover the case where NC U(2) is
broken by just a
fundamental Higgs; the Higgsac decouples.  We know from the calculations in
Sec. 3.3 that in this limit the theory is unitary.  To test the validity of
using the Higgsac to break the NC $U(1)$ subgroup we must see whether we
maintain a unitary theory for all values of $\beta$.

\begin{figure}
\begin{picture}(440,200)(10,10)
\Photon(80,190)(110,160){5}{3}
\Photon(80,130)(110,160){5}{3}
\Photon(110,160)(140,190){5}{3}
\Photon(110,160)(140,130){5}{3}
\put(62,172){$W^{+}_1$}
\put(62,145){$W^{-}_2$}
\put(135,172){$W^{+}_3$}
\put(135,145){$W^{-}_4$}

\Photon(190,185)(215,160){5}{3}
\Photon(190,130)(215,160){5}{3}
\Photon(215,160)(250,160){5}{3}
\Photon(250,160)(275,185){5}{3}
\Photon(250,160)(275,130){5}{3}
\put(172,170){$W^{+}_1$}
\put(172,142){$W^{-}_2$}
\put(278,170){$W^{+}_3$}
\put(278,142){$W^{-}_4$}
\put(218,170){$Z^{+},Z^{-}$}

\Photon(325,185)(395,185){5}{6}
\Photon(360,185)(360,130){5}{4}
\Photon(325,130)(395,130){5}{6}
\put(325,170){$W^{+}_1$}
\put(325,140){$W^{-}_2$}
\put(385,170){$W^{+}_3$}
\put(385,140){$W^{-}_4$}
\put(369,155.5){$Z^{+},Z^{-}$}

\Photon(130,75)(155,50){5}{3}
\Photon(130,20)(155,50){5}{3}
\Line(155,50)(190,50)
\Photon(190,50)(215,75){5}{3}
\Photon(190,50)(215,20){5}{3}
\put(112,60){$W^{+}_1$}
\put(112,32){$W^{-}_2$}
\put(218,60){$W^{+}_3$}
\put(218,32){$W^{-}_4$}
\put(170,55){$h$}

\Photon(265,75)(335,75){5}{6}
\Line(300,75)(300,20)
\Photon(265,20)(335,20){5}{6}
\put(265,60){$W^{+}_1$}
\put(265,30){$W^{-}_2$}
\put(325,60){$W^{+}_3$}
\put(325,30){$W^{-}_4$}
\put(308,46.5){$h$}

\end{picture}
\caption{Diagrams contributing to the scattering process $W^{+}_L W^{-}_L 
\rightarrow W^{+}_L W^{-}_L$ in NC $U(2)$ broken by both a fundamental Higgs
and Higgsac; the subscripts denote the 
momentum carried by the field.}
\label{u2fhw4diags}
\end{figure}
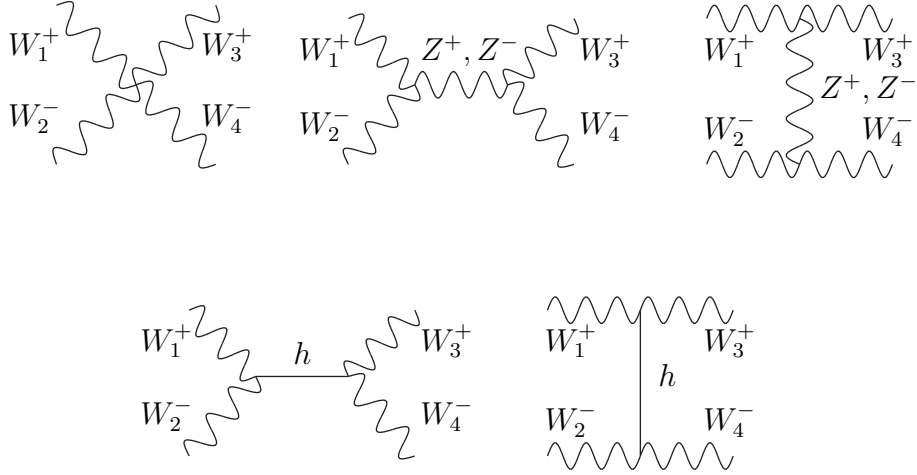

We again study the process $W^{+}_L W^{-}_L \rightarrow W^{+}_L W^{-}_L$.
This calculation requires the $W^{\pm}$ four-point coupling and the 
$h-W^{+}-W^{-}$ vertex, which are given Fig.~\ref{u2fafeyn}, as well as
the $Z^{+}-W^{+}-W^{-}$ and $Z^{-}-W^{+}-W^{-}$ vertices, which
are presented in Fig.~\ref{u2fhfeyn}.  The contributing diagrams are displayed
in Fig.~\ref{u2fhw4diags}.  Notice that since the Higgsac couples only to 
$Z^{\pm}$, it does not enter into this calculation; this is to be contrasted
with the model of Sec. 3.3, in which the addition of an adjoint Higgs
generated additional diagrams which were required to maintain unitarity in
$W^{+}_L W^{-}_L \rightarrow W^{+}_L W^{-}_L$.  Summing the contributions in
Fig.~\ref{u2fhw4diags}, we find that the ${\cal O}\left(s^2 
/m_{W}^4 \right)$ terms cancel as expected as these terms arise from the pure
gauge sector and are not affected by the details of the symmetry breaking.  
The ${\cal O}\left(s /m_{W}^2 \right)$ terms again split into those which are 
proportional to $c_{\theta}$ and those which are not, as in the previous 
examples.  These must cancel independently; for simplicity we list only 
the $c_\theta$ independent terms below:
\begin{eqnarray}
i{\cal M} &=& \frac{ig^2}{16} \frac{s}{m_{W}^2} \, \bigg\{ 6 \,{\rm exp} \left(
-ip_1 \wedge p_3 +ip_2 \wedge p_4 \right) -3\frac{m_{+}^2}{m_{W}^2} \bigg[ \,
c_{\alpha}^2 \, {\rm exp} \left(-ip_1 \wedge p_3 +ip_2 \wedge p_4 \right)
\nonumber \\ & & +s_{\alpha}^2 \, {\rm exp} \left(ip_1 \wedge p_3 -ip_2 
\wedge p_4 \right) -2c_{\alpha}s_{\alpha} 
\left( c_{(13)}c_{(24)}-s_{(13)}s_{(24)} 
\right) \bigg] \nonumber \\ & & -3\frac{m_{-}^2}{m_{W}^2} \bigg[ \,
c_{\alpha}^2 \, {\rm exp} \left(ip_1 \wedge p_3 -ip_2 \wedge p_4 \right)
+s_{\alpha}^2 \, {\rm exp} \left(-ip_1 \wedge p_3 +ip_2 \wedge p_4 \right)
\nonumber \\ & & 
+2c_{\alpha}s_{\alpha} \left( c_{(13)}c_{(24)}-s_{(13)}s_{(24)} \right) \bigg] 
\bigg\}+ c_{\theta} \left[ \ldots \right] \,\, .
\label{u2fhw4amp}
\end{eqnarray}
In obtaining this expression we have used the relations 
in Eq.~\ref{ids1}.  It is clear that this 
expression does not vanish for general $\beta$.  As an example, let us
examine an arbitrary 
case with ${\rm tan}\left( \beta \right) =\sqrt{3}$, $s_{\alpha} =
1/2$, $c_{\alpha}=\sqrt{3}/2$; with this choice of parameters the amplitude
takes the very simple form
\begin{equation}
i{\cal M}= 12\sqrt{3} \, s_{(13)}s_{(24)} 
+ c_{\theta} \left[ \ldots \right] \,\, .
\label{u2fhw4amp2}
\end{equation}
This clearly vanishes in the commutative 
limit when $U(2)$ splits into the product of
independent group factors $SU(2) \times U(1)$, but not in a general NC setting.
In addition, we have found that a
similar lack of cancellation of ${\cal O}\left(s/m_{-}^2
\right)$ terms occurs in the process $hZ^{-}_L \rightarrow hZ^{-}_L$. Since 
the difference between the previous test case, where unitarity was maintained,
and the one presented here was to
trade an adjoint Higgs for a Higgsac in the symmetry breaking, we thus 
conclude that the use of the Higgsac representation in NC symmetry 
breaking induces unitarity violations in processes involving
longitudinal gauge bosons. 

As can be seen from the above discussion, our survey of toy models has allowed 
us to isolate the source of the tree-level unitarity breakdown in the 
proposed NCSM: the 
breaking of the symmetry by the Higgsac fields. 

\begin{figure}
\begin{picture}(200,190)(10,10)

\Photon(10,50)(50,50){5}{3}
\put(10,59.72){$Z^{-}_{\rho}$}
\put(10,36.72){$\leftarrow$  $p_3$}
\Photon(50,50)(78.28,78.28){5}{3}
\put(64,55){$\nearrow$ $p_1$}
\put(82,70){$W^{-}_{\mu}$}
\Photon(50,50)(78.28,21.72){5}{3}
\put(80,19.72){$W^{+}_{\nu}$}
\put(64,40.72){$\searrow$ $p_2$}
\put(100,50){$= \frac{ig}{\sqrt{2}} \left( c_{\alpha} \, e^{ip_1 \wedge p_2}
+s_{\alpha} \, e^{-ip_1 \wedge p_2} \right) \left[ (p_1 -p_2)_{\rho} 
g_{\mu\nu} +(p_2 -p_3)_{\mu}g_{\nu \rho} +(p_3 -p_1)_{\nu}g_{\mu\rho}\right]$}

\Photon(10,148.28)(50,148.28){5}{3}
\put(10,158){$Z^{+}_{\rho}$}
\put(10,135){$\leftarrow$  $p_3$}
\Photon(50,148.28)(78.28,176.56){5}{3}
\put(64,153.28){$\nearrow$ $p_1$}
\put(82,168.28){$W^{-}_{\mu}$}
\Photon(50,148.28)(78.28,120){5}{3}
\put(80,118){$W^{+}_{\nu}$}
\put(64,138){$\searrow$ $p_2$}
\put(100,148.28){$= -\frac{ig}{\sqrt{2}} \left( c_{\alpha} \, e^{-ip_1 \wedge 
p_2} -s_{\alpha} \, e^{ip_1 \wedge p_2} \right)\left[ (p_1 -p_2)_{\rho} 
g_{\mu\nu} +(p_2 -p_3)_{\mu}g_{\nu \rho} +(p_3 -p_1)_{\nu}g_{\mu\rho}\right]$}

\end{picture}
\caption{Additional Feynman rules necessary for $W^{+}_L W^{-}_L \rightarrow
W^{+}_L W^{-}_L$ in NC $U(2)$ broken by both a fundamental Higgs and Higgsac.}
\label{u2fhfeyn}
\end{figure}

\vspace*{-1.0cm}

\vspace{1.0cm}
\section{Discussion and Conclusion}

The straightforward construction of a NC version of the SM is made difficult 
by the requirements of gauge invariance and the constraints that arise from 
group theory. Once these conditions are met we expect any realistic 
perturbative NC model that reduces to the SM in the commutative limit to 
possess a number of essential properties in order to be a viable theory
and make phenomenological 
predictions: renormalizability, anomaly freedom, tree-level 
unitarity, \etc.  In this paper we have explicitly shown that the version of 
the NCSM constructed by Chaichian \etal{\cite {Chaichian:2001py}} does not 
satisfy the requirement of tree-level unitarity for gauge boson scattering at
high energies.  In order to track down the origin of this 
unitarity violation we have examined a number of simpler NC gauge 
models with various gauge structures wherein the symmetry was spontaneously 
broken by various Higgs representations. It
was clear from this study that the use of the Higgsac representations to 
break the gauge symmetries is the source of this unitarity violation. 
In the SM the mass relationship between the $W^\pm$ and $Z$ 
fields as well as the Higgs exchanges ensures cancellations
of the leading and sub-leading unitarity violating terms for this process. 
In the version of the NCSM discussed here there are no relationships between 
the new gauge boson masses themselves or with those of the $W^\pm$ 
and $Z$ and there are no Higgsac contributions 
to induce cancellations. Since these 
$m_{G^{0}}^2$ and $m_{W^{0}}^2$ terms are unavoidable in this approach 
unitarity must fail and this version of the NCSM becomes phenomenologically 
unacceptable. 

If Higgsac representations cannot be used in the symmetry breaking of the 
NCSM then we are faced with a severe model construction problem. The use of 
products of $U(n)$ groups, required by gauge invariance and algebraic
closure in the Moyal approach, necessitates the 
breaking of their $U(1)$ subgroups as the first stage of symmetry breaking 
since they cannot be identified with the $U(1)_Y$ gauge symmetry of the SM. 
As noted earlier, ordinary Higgs representations such as fundamentals or 
adjoints will not break these $U(1)$'s but will instead break the $SU(n)$ 
subgroups; this must be avoided.  Using larger NC gauge groups does not
help the situation as fundamental and adjoint vevs only induce the breaking
pattern $U(n)\to U(n-1)$ instead of breaking to $SU(n-1)$.

We thus conclude that it is difficult, if not impossible, to build a
phenomenologically viable non-commutative Standard Model under the
Weyl-Moyal approach.

\noindent{\Large\bf Acknowledgements}

\noindent
We would like to thank Mohammad Sheikh-Jabbari for valuable discussions.
The work of F. P. was supported in part by the NSF Graduate Research Program.

\newpage

%
\def\MPL #1 #2 #3 {Mod. Phys. Lett. {\bf#1},\ #2 (#3)}
\def\NPB #1 #2 #3 {Nucl. Phys. {\bf#1},\ #2 (#3)}
\def\PLB #1 #2 #3 {Phys. Lett. {\bf#1},\ #2 (#3)}
\def\PR #1 #2 #3 {Phys. Rep. {\bf#1},\ #2 (#3)}
\def\PRD #1 #2 #3 {Phys. Rev. {\bf#1},\ #2 (#3)}
\def\PRL #1 #2 #3 {Phys. Rev. Lett. {\bf#1},\ #2 (#3)}
\def\RMP #1 #2 #3 {Rev. Mod. Phys. {\bf#1},\ #2 (#3)}
\def\NIM #1 #2 #3 {Nuc. Inst. Meth. {\bf#1},\ #2 (#3)}
\def\ZPC #1 #2 #3 {Z. Phys. {\bf#1},\ #2 (#3)}
\def\EJPC #1 #2 #3 {E. Phys. J. {\bf#1},\ #2 (#3)}
\def\IJMP #1 #2 #3 {Int. J. Mod. Phys. {\bf#1},\ #2 (#3)}
\def\JHEP #1 #2 #3 {J. High En. Phys. {\bf#1},\ #2 (#3)}

\end{document}